# Phosphorene nanoribbons, nanotubes and van der Waals multilayers

Hongyan Guo[1,2,⊥], Ning Lu[3,2,⊥], Jun Dai[2], Xiaojun Wu[1,4,5]*, Xiao Cheng Zeng[2,4],*

[1]CAS Key Lab of Materials for Energy Conversion, Department of Materials Science and Engineering, University of Science and Technology of China, Hefei, Anhui 230026, China, [2]Department of Chemistry and Department Mechanics and Materials Engineering, University of Nebraska-Lincoln, Lincoln, NE 68588, USA, [3]Center for Nano Science and Technology, Department of Physics, Anhui Normal University, Wuhu, Anhui 241000, China, [4]Hefei National Laboratory for Physical Sciences at the Microscale, University of Science and Technology of China, Hefei, Anhui 230026, China, [5]Synergetic Innovation Center of Quantum Information & Quantum Physics, University of Science and Technology of China, Hefei, Anhui 230026, China
*Email: xzeng1@unl.edu; xjwu@ustc.edu.cn

## ABSTRACT

We perform a comprehensive first-principles study of the electronic properties of phosphorene nanoribbons, phosphorene nanotubes, multilayer phosphorene, and heterobilayers of phosphorene and two-dimensional (2D) transition metal dichalcogenide (TMDC) monolayer. The tensile strain and electric-field effects on electronic properties of low-dimensional phosphorene nanostructures are also investigated. Our calculations show that zigzag phosphorene nanoribbons (z-PNRs) are metals, regardless of the ribbon width while armchair phosphorene nanoribbons (a-PNRs) are semiconductors with indirect bandgaps and the bandgaps are insensitive to variation of the ribbon width. We find that tensile compression (or expansion) strains can reduce (or increase) the bandgap of the a-PNRs while an in-plane electric field can significantly reduce the bandgap of a-PNRs, leading to the semiconductor-to-metal transition beyond certain electric field. For single-walled phosphorene nanotubes (SW-PNTs), both armchair and zigzag nanotubes are semiconductors with direct bandgaps. With either tensile strains or transverse electric field, similar behavior of bandgap modulation can arise as that for a-PNRs. It is known that multilayer phosphorene sheets are semiconductors with their bandgaps decreasing with increasing the number of multilayers. In the presence of a vertical electric field, the bandgaps of multilayer phosphorene sheets decrease with increasing the electric field, and the bandgap modulation is more significant with more layers. Lastly, heterobilayers of phosporene with a TMDC ($MoS_2$ or $WS_2$) monolayer are still semiconductors while their bandgaps can be reduced by applying a vertical electric field as well.

**INTRODUCTION**

Two-dimensional (2D) materials with atomic thickness, such as carbon graphene, boron-nitride, and 2D transition metal dichalcogenides (TMDCs), have aroused great interests due to their unique properties not seen in their bulk counterparts.[1-4] Graphene is known to have some remarkable electronic and mechanical properties such as high carrier mobility but the absence of a bandgap limits its performance in providing relatively large off current and high on-off ratio. As a representative of 2D TMDCs, $MoS_2$ monolayer possesses a direct bandgap of ~1.8 eV and relatively high on/off ratio. However, the carrier mobility of $MoS_2$ is just several tens of $cm^2V^{-1}s^{-1}$, much lower than that of graphene.[5,6]

Recently, a new 2D material, namely, layered black phosphorus or phosphorene,[7] has been isolated in the laboratory through mechanical exfoliation from bulk black phosphorus and has immediately received considerable attention.[8-17] It turns out that phosphorene possess some remarkable electronic properties as well. For example, it is reported that phosphorene has the drain current modulation up to $10^5$ and carrier mobility up to 1000 $cm^2V^{-1}s^{-1}$, which makes phosphorene a potential candidate for future nanoelectronic applications.[12] Phosphorene also has a direct bandgap which can be modified from 1.51 eV for a monolayer to 0.59 eV for a five-layer.[14] Moreover, the *p*-type black phosphorene transistor has already been integrated with the *n*-type $MoS_2$ transistor to make a 2D CMOS inverter.[13]

Tunability of electronic properties of 2D materials is crucial for their applications in optoelectronics. Several strategies are commonly used: (1) Converting a 2D sheet to 1D structure, such as 1D nanoribbons or nanotubes.[18-20] (2) Varying the number of stacked 2D sheets of the same material or constructing van der Waals heterolayers by stacking different 2D materials.[20-26] (3) Applying either an external electric field (in-plane or vertical) or a tensile strain, which is widely used for tuning bandgaps of 2D or 1D materials.[22-24, 27-38] A strain may even induce the semiconductor-to-metal transition; and strain controlled anisotropic conductance has also been predicted for phosphorene.[10, 15]

In this article, we report a systematic study of electronic properties of phosphorene nanoribbons (PNRs), single-walled phosphorus nanotubes (SW-PNTs), multilayer phosphorene, and heterobilayers with a 2D TMDC sheet. Effects of external electric field or tensile strain have also been investigated. Our study suggests that low-dimensional phosphorene-based nanostructures are very versatile and either a mechanical or an electric pathway can be adopted to engineer their bandgaps and other electronic properties.

**RESULTS AND DISCUSSION**

**1. One-dimentioanl phosphorene nanoribbons**

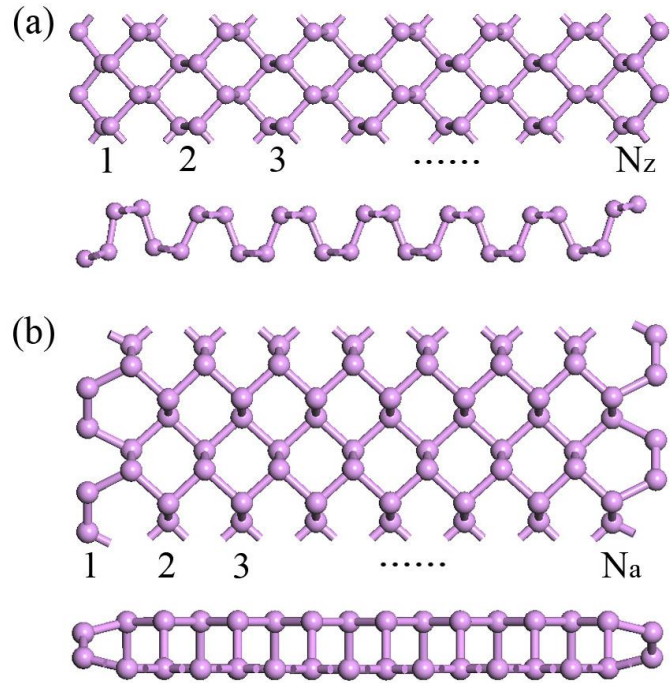

**Figure 1.** Top and side views of structures of phosphorene nanoribbons (PNRs): (a) a zigzag PNR (z-PNR) and (b) an armchair PNR (a-PNR).

Figure 1 illustrates the structures of PNRs with either two zigzag or two armchair egdes (where the periodicity is along the ribbon direction). Electronic structures of z-PNRs with width $N_z$ = 7-12 are computed, and all exhibit metallic character (see Supporting Information Figure S1). The computed electronic structure of 8-zigzag PNR is shown in Figure 2a, where the partial charge density analysis shows that the metallic character stems from the P atoms near the edges. When an in-plane electric field $E_{ext}$ in the range of 0.1 ~ 0.5 V/Å is applied across zigzag PNRs (*i.e.*, normal to the ribbon direction), the z-PNRs remain to be metallic. We have also investigated effect of tensile strains on the electronic properties of z-PNRs, and found that the strains have little effect on the band structures of z-PNRs. Additional computation based on the HSE06 functional confirms that the z-PNRs remain to be metallic under the tensile strain.

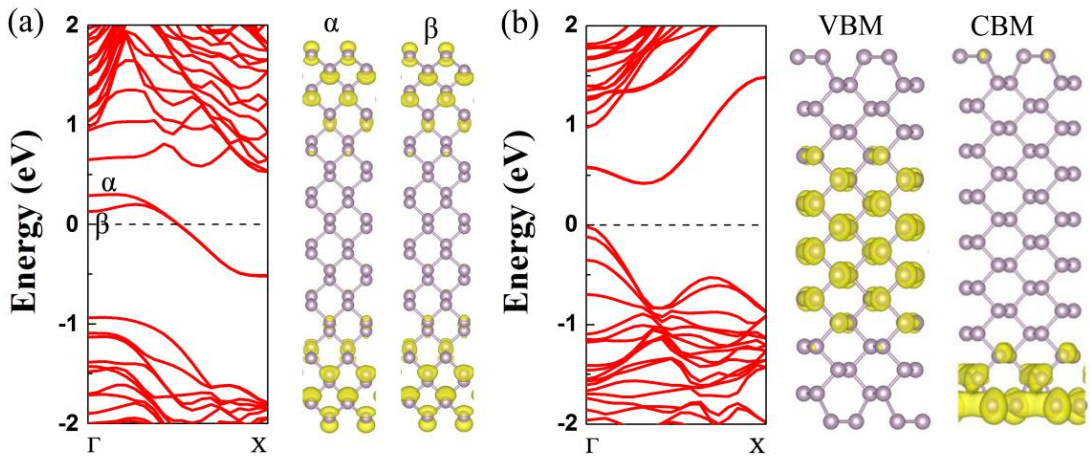

**Figure 2**. Computed band structures (PBE) of (a) 8-zigzag and (b) 8-armchair PNRs. Right panels in (a) show the partial charge density distribution of the corresponding α and β bands, respectively. Right panels in (b) show the partial charge density distribution of the valence band maximum (VBM) and the conduction band minimum (CBM), respectively. The isosurface value is 0.002 e/Bohr$^3$.

Contrary to metallic z-PNRs, all armchair PNRs are semiconductors with indirect bandgaps. For a-PNRs with width $N_a$ = 7-12, their bandgaps exhibit little variation with the width (see Figure S2). A 7-armchair PNR has an indirect bandgap of 0.44 eV (PBE calculation), while a 9-archiar PNR has an indirect bandgap of 0.42 eV. Computed electronic structure of 8-armchair PNR is shown in Figure 2b where the partial charge density analysis indicates that the VBM is contributed by the P atoms in central region of the PNR while the CBM is contributed by the P atoms in one edge of the 8-armchair PNR.

Unlike z-PNRs, the tensile strain has significant effect on the electronic properties of a-PNRs. As shown in Figure 3a, the bandgap of a 8-armchair PNR exhibits a linear response to the tensile strain, ranging from being 0.31 eV at -4% compression strain to being 0.52 eV at 4% expansion strain. The HSE06 results also confirm the same trend on strain-dependent bandgap. The computed band structures corresponding to the -4% and 4% strains are plotted in Figure S3. One can see that the shift of VBM is a main reason for the bandgap modification. The VBM is mainly contributed by P-P bonding states. When the P-P bond length is elongated for about 0.02 Å from the -4% to 4% strain, the VBM is shifted by 0.17 eV. On the other hand, the CBM is mainly contributed by the edge atoms and is thus sensitive to the reconstruction of edge atoms. Hence, the tensile strain has much less effect on the CBM.

As shown in Figure 3b, the armchair PNRs also exhibit significant response to the in-plane transverse electric field. Taking 8-armchair PNR as an example, its bandgap is 0.43 eV. Under the in-plane transverse electric field of 0.1 V/Å, the bandgap is reduced to 0.10 eV. When magnitude of the electric field increases to 0.2 V/Å, the bandgap vanishes so that the 8-armchair PNR undergoes a semiconductor-to-metal transition. Further increasing the electric field does not change the metallic character of the a-PNRs. The HSE06 results are consistent with PBE results, except that the semicondutor-metal transition occurs at 0.4 V/Å. The bandgap reduction with increasing the electric field can be understood by the Stark effect.[20] Notably, charge of the PNR redistributes under the electric field, as shown in Figure S4. In particular, charge corresponding to CBM and VBM changes in opposite direction with changing the electric field, leading to the downshift of CBM and upshift of VBM.

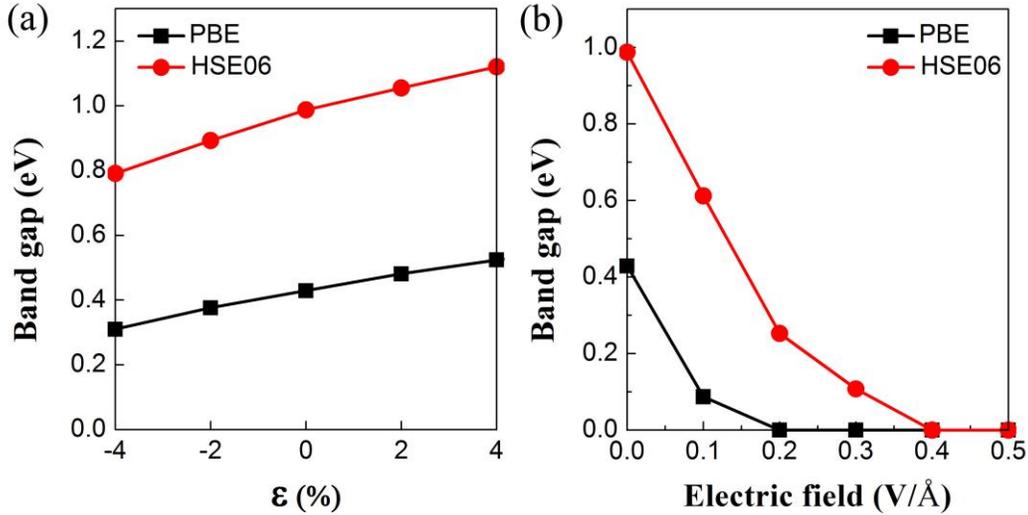

**Figure 3**. (a) Computed band gaps of 8-armchair PNR versus the tensile strain, ranging from –4% to 4%. Here, "-" represents compression and "+" represents expansion. (b) Computed band gaps versus in-plane transverse electric field for the 8-armchair PNR.

To examine relative stabilities PNRs, we compute the cohesive energy per atom for both zigzag and armchair PNRs versus the ribbon width. Here, the cohesive energy $E_c$ is defined as $E_c = nE_p - E_{Pn}$, where $E_p$ is the total energy of a single P atom, $E_{Pn}$ is the total energy of a PNR, and $n$ is the number of P atoms in the PNR supercell. PNRs with greater cohesive energy per atom are more stable. As shown in Figure S5, the cohesive energy of armchair PNRs increases gradually with increasing the ribbon width. Clearly, the cohesive energy of zigzag PNRs are greater than that of armchair PNRs with the same ribbon width. For example, the cohesive energy of 7-zigzag PNR is 3.64 eV, greater than that (3.60 eV) of 7-armchair PNR. As a comparison, the computed cohesive energy per atom of bulk black phosphorus is 3.79 eV.

## 2. One-dimentioanl single-walled phosphorene nanotubes

As shown in Figure 4, a SW-PNT can be viewed as rolling up a phosphorene monolayer. A previous study reported a SW-PNT constructed by rolling up a flat phosphorus sheet with graphene-like structure.[40] Here, a monolayer phosphorene with ridged structure is directly rolled up to construct the SW-PNT. The conformation of any specific SW-PNT can be described with a pair of integer indexes ($n_1$, $n_2$) that defines a rollup vector R = $n_1a_1 + n_2a_2$ (see Figure 4). Two types of SW-PNTs are considered in this study, zigzag ($n_1$, 0) and armchair (0, $n_2$). A number of small diameter SW-PNTs with diameter ranging from 12 to 19 Å are investigated.

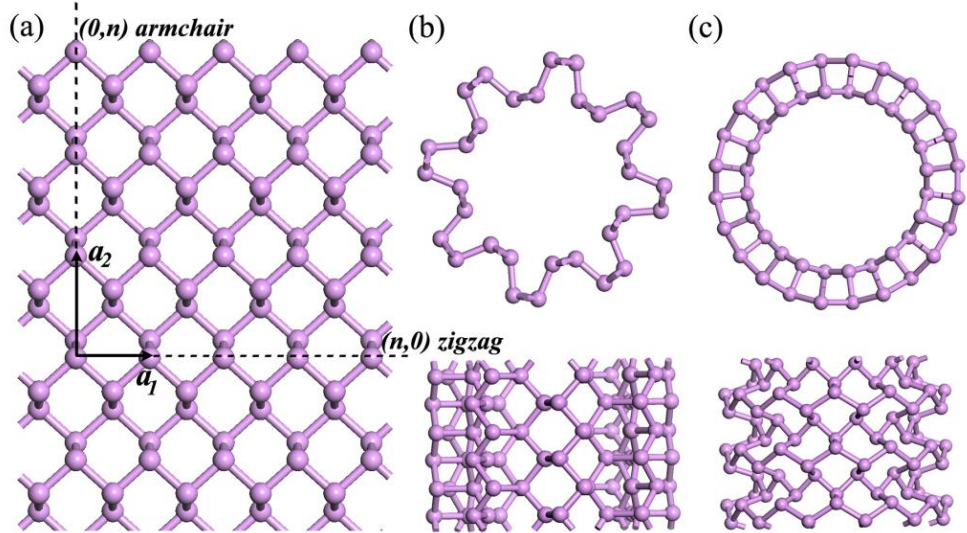

**Figure 4.** Schemtic plots of SW-PNTs which can be viewed as rolling up a phosphorene sheet following the roll-up vector R = $n_1a_1 + n_2a_2$. Top and side views of a structure of (b) armchair and (c) zigzag SW-PNTs.

As shown in Table 1, for armchair PNTs, the optimized unit-cell length $L_c$ in the axial direction is 3.30 Å, independent to the diameter of a-PNTs. However, for zigzag PNTs, the optimized unit-cell length $L_c$ gradually increases with $D$, from $L_c$ = 4.13 Å for (12, 0) to $L_c$ = 4.48 Å for (17, 0). The calculated strain energy per atom, defined as the cohesive energy difference between the PNT and a perfect phosphorene monolayer, is also shown in Table 1. The strain energy per atom can be used to evaluate relative stabilities of the PNTs. For both armchair and ziazag PNTs, as expected, the strain energy decreases with increasing $D$. However, the armchair PNTs give rise to much smaller strain energies compared to the zigzag PNTs with nearly the same $D$, indicating that the a-PNTs are energetically more favorable than z-PNTs.

Table 1. Diameter of SW-PNTs $D$ (in Å), the unit-cell length $L_c$ in the axial direction (in Å), the computed bandgap $E_{g1}$ (in eV) based on PBE functional, the bandgap $E_{g2}$ (in eV) based on HSE06 functional, and the strain energy $E_S$ (in eV/atom).

| SW-PNT | $D$ | $L_c$ | $E_{g1}$ | $E_{g2}$ | $E_S$ |
|---|---|---|---|---|---|
| (0, 8) | 12.50 | 3.30 | 0.28 | 0.76 | 0.06 |
| (0, 9) | 13.86 | 3.30 | 0.36 | 0.91 | 0.04 |
| (0, 10) | 15.24 | 3.30 | 0.44 | 1.01 | 0.03 |
| (0, 11) | 16.66 | 3.30 | 0.51 | 1.10 | 0.02 |
| (0, 12) | 18.12 | 3.30 | 0.56 | 1.16 | 0.02 |
| (0, 13) | 19.48 | 3.30 | 0.61 | 1.22 | 0.01 |
| (12, 0) | 14.28 | 4.13 | 0.10 | 0.47 | 0.23 |
| (13, 0) | 15.23 | 4.17 | 0.13 | 0.57 | 0.20 |
| (14, 0) | 16.25 | 4.20 | 0.14 | 0.62 | 0.18 |
| (15, 0) | 17.16 | 4.30 | 0.22 | 0.74 | 0.16 |
| (16, 0) | 18.10 | 4.40 | 0.31 | 0.92 | 0.14 |
| (17, 0) | 18.98 | 4.48 | 0.35 | 0.99 | 0.13 |

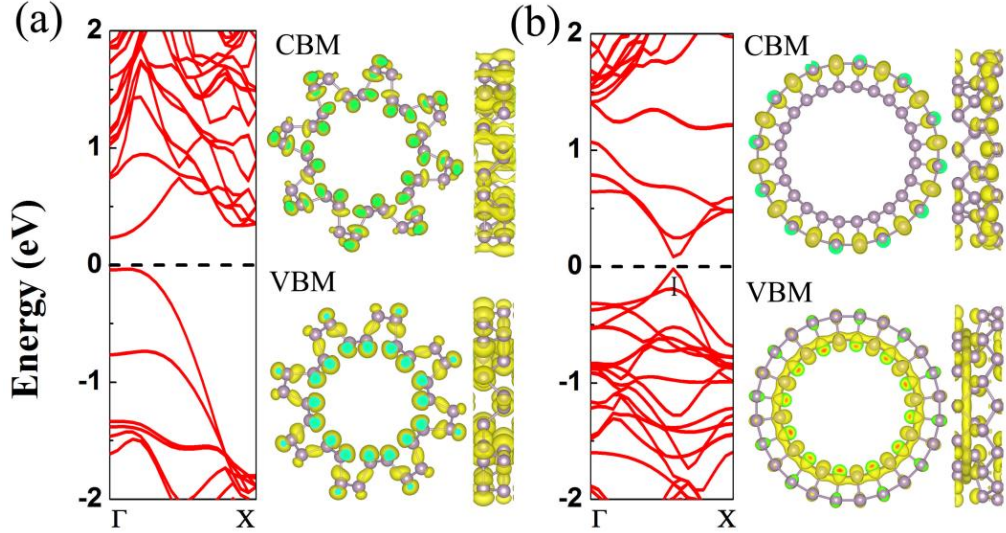

**Figure 5**. Computed band structure (PBE) and corresponding partial charge density distribution of (a) armchair (0, 8) and (b) zigzag (12, 0) SW-PNTs. The isosurface value is 0.001 e/ Bohr$^3$.

Importantly, both a-PNTs and z-PNTs are semiconductors with their bandgaps increase with $D$. For armchair SW-PNTs, the bandgap increases from 0.28 eV (0.76 eV based on HSE06) for $D$ = 12.50 Å to 0.61 eV (1.22 eV based on HSE06) for $D$ = 19.48 Å. For zigzag SW-PNTs, the bandgap increases from 0.1 eV (0.47 eV based on HSE06) for $D$ = 14.28 Å to 0.3 eV (0.99 eV based on HSE06) for $D$ = 18.98 Å. The computed electronic structures of armchair (0, 8) and zigzag (12, 0) SW-PNTs are shown in Figure 5. The armchair (0, 8) SW-PNT has a direct bandgap with its VBM and CBM being located at the Γ point. As shown in Figure 5a, the VBM manifests a bonding character between the inner and outer P atoms, while the CBM exhibits a diffusive state of the outer or inner P-P bond along the axis. The zigzag (12, 0) SW-PNT has a direct bandgap with the CBM and VBM being located at the I point. The VBM manifests mainly the bonding state between inner atoms, while the CBM is mainly due to isolated $p_z$ states of outer atoms.

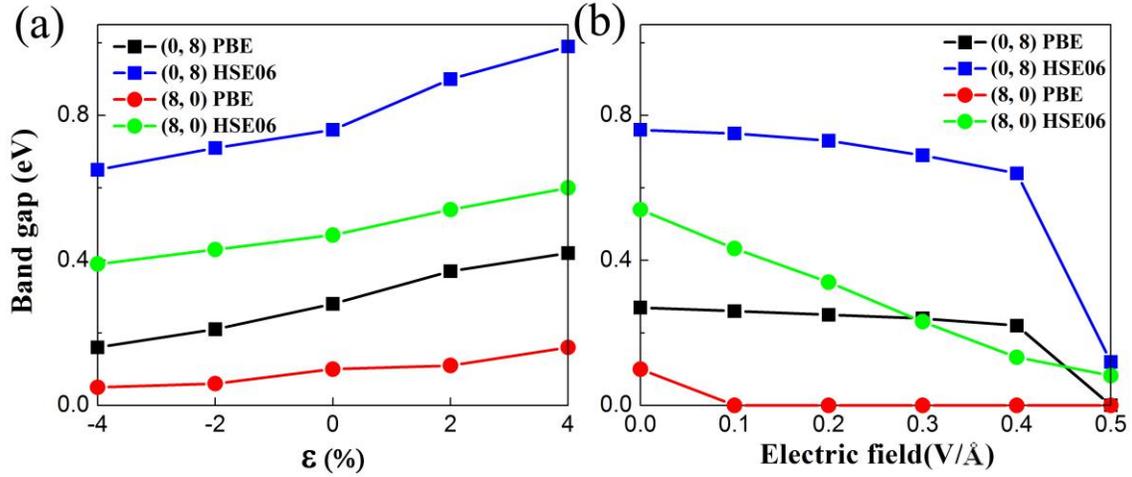

**Figure 6**. Computed bandgaps (PBE and HSE06) of armchair (0, 8) and zigzag (12, 0) SW-PNTs versus (a) tensile strain, ranging from –4% to 4%, where "-" represents compression and "+" indicates expansion; and (b) the transverse electric field, ranging from 0.1 to 0.5 V/Å.

Again, electronic properties of the SW-PNTs can be modified through either elongation or compression of the nanotubes in the axial direction. For armchair (0, 8) SW-PNT, it undergoes an direct-to-indirect bandgap transistion under a critical tensile strain (Figure S6). The CBM is still located at the $\Gamma$ point, while the VBM is shifted along the $\Gamma$-X line. As shown in Figure 6a, the bandgap increases up to 0.42 eV (0.99 eV for HSE06) with increasing the expansion strain up to 4%. The bandgap decreases gradully with the compression strain down to -4%, where the corresponding bandgap is 0.17 eV (or 0.65 eV based on HSE06). The PBE results indicate that the bandgap is enlarged by ~0.25 eV with the strain changing from -4% to 4%, while the HSE06 results show the same trend with the bandgap enlarged by 0.34 eV. As shown in Figure S6, the upshift of the CBM for about 0.19 eV with the strain changing from -4% to 4% is the main reason for the bandgap enlargement. Since the CBM is mainly contributed by the outer P-P bond along the axial direction, any appreciable deformation along the axial direction would significantly affect bandgaps of armchair SW-PNTs. For zigzag (12, 0) SW-PNT, its bandgap also increases with increasing the tensile strain. However, both the CBM and VBM are shifted downward under the strain while the VBM is shifted downward more than the CBM, thereby leading to net increase of the bandgap (Figure S7).

Figure 6b shows that when the armchair (0, 8) PNT is subjected to a transverse electric field, its bandgap decreases with increasing the field strength. A semiconductor-to-metal transition occurs when the electric field reaches 0.5 V/Å. For zigzag (8, 0) PNT, the bandgap also decreases with increasing the field strength, while the semiconductor-to-metal transition occurs when the field reaches 0.1 V/Å. Moreoever, HSE06 results confirm the same trend on the field-dependent bandgap. Again, the gap modulation can be understood through the Stark effect. Overall, our calculations show that either the tensile strain or transverse electric field can significantly affect bandgaps of PNTs, regardless of chiralty of the PNTs.

**3. Multilayer phosphorene sheets**

It is known that the bulk black phosphorus exhibits the AB stacking in packed layers. Hence, we also consider the AB-stacking for multilayer phosphorene sheets (see Figure S8). The optimized cell parameters and interlayer distance for 2 - 4 layer phosphorene sheets are listed in Table S1, which are found in agreement with previous studies.[14] Computed electronic structures of 1 - 4 layer phosphorene sheets, based on the PBE calculations, are shown in Figure 7. The phosphorene monolayer is a semiconductor with a direct bandgap of 0.84 eV at the Γ point (Figure 7a), consistent with previous results.[39] Bilayer, trilayer and 4-layer phosphorene sheets are also semiconductors with a direct bandgap of 0.44, 0.25 and 0.16 eV, respectively. The bandgaps of multilayer phosphorene sheets decrease with increasing the number of layers. Calculations based on the HSE06 functional also confirm this trend (see Figure S9), giving direct bandgap of 1.54, 1.11, 0.91 and 0.82 eV for monolayer, bilayer, trilayer and 4-layer phosphorene sheets, respectively.

As shown in Figure 8a, when a phosphorene monolayer is subjected to a vertical electric field, its bandgap exhibits little change. However, for a phosphorene bilayer, its bandgap decreases slightly upon application of the verticle electric field. With increaseing the number of layers, the bandgap reduction due to the vertical electric field becomes more significant. Such a layer-dependent bandgap reduction due to the vertical electric field can be understood by the Stark effect. The difference in electric potential among different phophorene layers can be induced by the vertical electric field.[23] As a result, the energy bands due to different phosphorene layers are further split by the vertical electric field. The splitting is proportional to the number of layers as well as the strength of the vertical electric field, thereby resulting in stronger bandgap modulation for thicker phosphorene sheets.

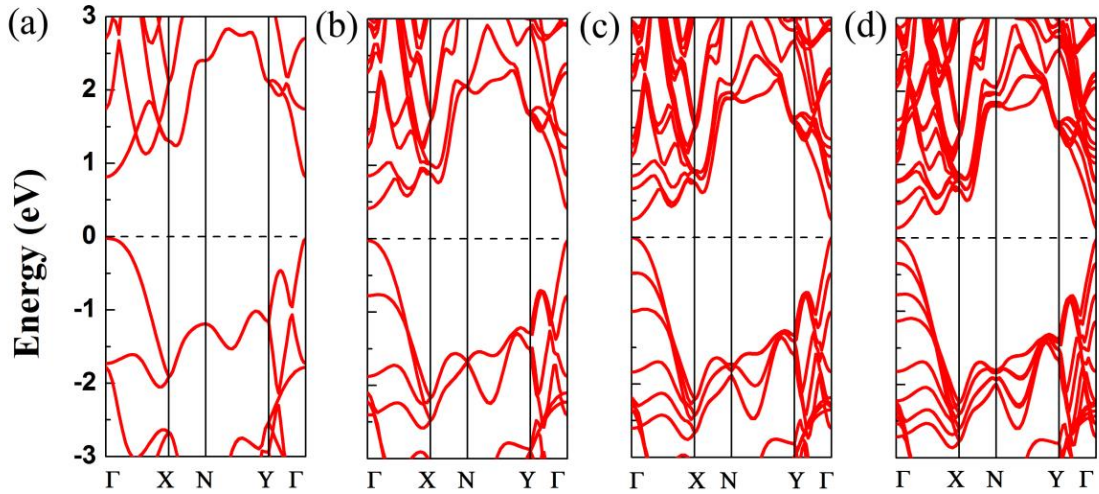

**Figure 7**. Computed band structures ( PBE) of (a) monolayer, (b) bilayer, (c) trilayer, and (d) 4-layer phosphorene sheets.

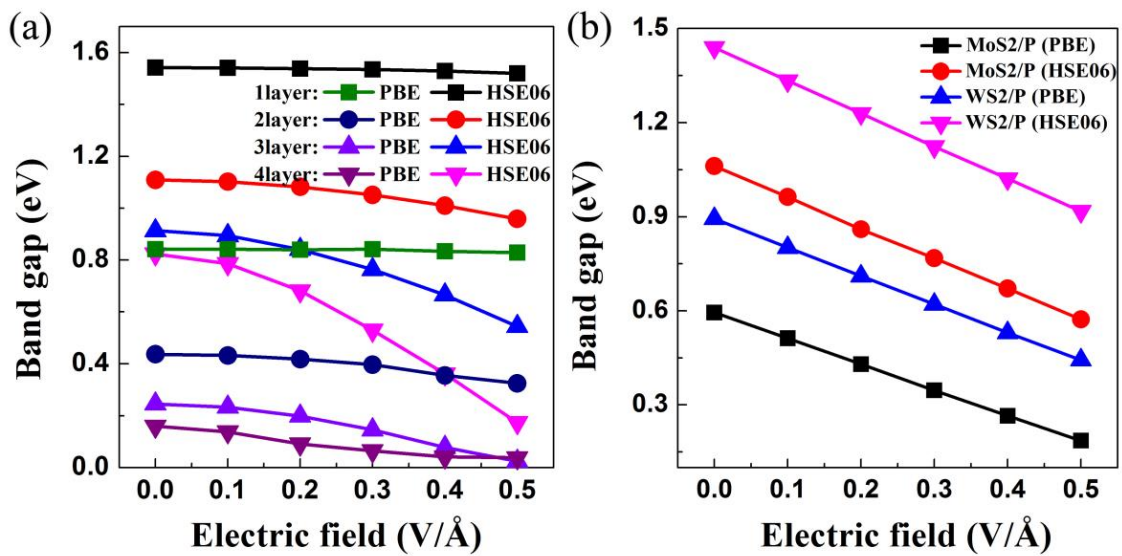

**Figure 8**. Vertical electric field dependent bandgap (PBE and HSE06) for (a) multilayer phosphorene sheets, and (b) TMDC/phosphorene heterobilayers.

## 4. Heterobilayers of phosphorene and TMDC monolayer

A recent experiment demonstrated successful preparation of a CMOS inverter which combines a phosphorene PMOS transistor with a $MoS_2$ NMOS transistor. Here, electronic properties of $MoS_2$/phosphorene and $WS_2$/phosphorene heterostructure are computed, as shown in Figure 9.

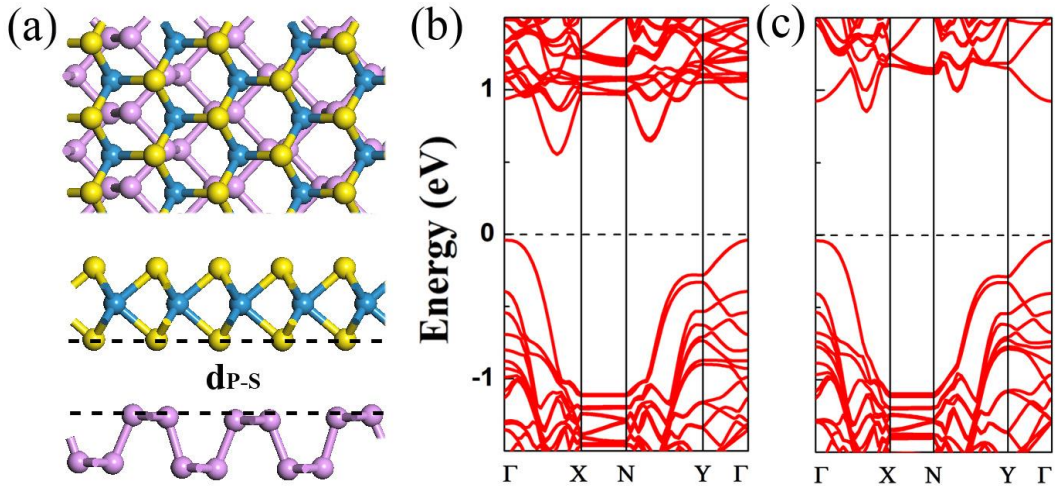

**Figure 9**. (a) Top and side views of the optimized $MoS_2$ (or $WS_2$)/phosphorene heterobilayer structure. (b) Computed band structures (PBE) of (b) $MoS_2$/phosphorene and (c) $WS_2$/phosphorene heterobilayers. Blue, yellow and purple spheres represent Mo (or W), S and P atoms, respectively. The $d_{P-S}$ denotes the interlayer height difference between P and S atoms.

The distance between $WS_2$ and phosphorene monolayer is 3.30 Å and the distance between $MoS_2$ and phosphorene monolayer is 3.21 Å, suggesting weak van der Waals interaction between $MoS_2$ ($WS_2$) and phosphorene layers. We also compute the binding energy of between the two layers. Here, the binding energy is defined as: $E_{BE} = (E_P + E_{MoS2(WS2)} - E_{MoS2(WS2)/P})/n$, where $E_P$, $E_{MoS2(WS2)}$, and $E_{MoS2(WS2)/P}$ are the total energy of a phosphorene monolayer, a $MoS_2$ ($WS_2$) monolayer and a $MoS_2(WS_2)$/phosphorene heterobilayer, respectively; $n$ is the total number of atoms per supercell. The computed binding energies of $MoS_2$/phosphorene and $WS_2$/phosphorene heterobilayers are 36 and 35 meV per atom, respectively, again consistent with weak van der Waals interaction between $MoS_2$ ($WS_2$) and phosphorene layers.

Figure 9b,c show that both $MoS_2$/phosphorene and $WS_2$/phosphorene heterobilayers are semiconductors with an indirect bandgap of 0.59 and 0.89 eV (PBE), respectively. The partial density of state (PDOS) analysis suggests that the VBM is mainly contributed by the *p* orbital of P atoms and the CBM is mainly contributed by the *d* orbital of Mo (or W) atoms (Figure S10). HSE06 calculations give indirect bandgaps of 1.06 and 1.44 eV for $MoS_2$/phosphorene and $WS_2$/phosphorene heterobilayers, respectively. A previous theoretical study has shown that the bandgap of TMDC heterobilayers can be tuned by a vertical electric field.[25] Likewise, as shown in Figure 8b, a vertical electric field can significantly reduce the bandgap of TMDC/phosporene heterobilayers as well.

**CONCLUSIONS**

In conclusion, we have performed a first-principles study of structural and electronic properties of PNRs, SW-PNTs, multilayers, and 2D TMDC/phosphorene heterobilayers, as well as tunability of their bandgaps by applying either a tensile strain or an external electrical field. Our calculations show that the zigzag PNRs are metallic, while the armchair PNRs are semiconducting with indirect bandgaps. All armchair and zigzag SW-PNTs are semiconducting with direct bandgaps. The bandgap of armchair PNRs and SW-PNTs can be modulated by either the tensile strain or transverse electric field. All multi-layer phosphorene sheets are semiconduting with their bandgap decreasing with more stacked layers. Moreover, a vertical electric field can modulate bandgaps of multilayer phosphorene sheets, as well as those of 2D TMDC/phosphorene heterobilayers. These novel electronic properties of low-dimensional (1D and 2D) phosphorene nanostructures, in conjuction with their remarkable strain enegineering and electric field effects in tuning the bandgaps, suggest that phosphorene is a promising candidate for future nanoelectronic and optoelectronic applications.

## COMPUTATIONAL DETAILS

In this work, all calculations are performed within the framework of spin-polarized density functional theory (DFT), implemented in the Vienna ab initio simulation package (VASP).[41,42] The generalized gradient approximation (GGA) in the form of the Perdew-Burke-Ernzerhof (PBE) functional, and projector augmented wave (PAW) potentials are used.[43-45] Effect of van der Waals (vdW) interaction is accounted for by using a dispersion corrected DFT method (optB88-vdW functional),[46, 47] which has been proven reliable for multilayer phosphorene systems.[14] More specifically, a $1 \times 1$ unit cell is adopted for all calculations except for $MoS_2$/phosphorene and $WS_2$/phosphorene heterobilayers for which the $1\times5$ $MoS_2$ or $WS_2$ supercell that nearly matches the $1\times6$ phosphorene supercell (with the lattice mismatch less than 5%) is adopted. The vacuum size is larger than 15 Å between two adjacent images. An energy cutoff of 500 eV is adopted for the plane-wave expansion of the electronic wave function. Geometry structures are relaxed until the force on each atom is less than 0.01 eV/Å, and the energy convergence criteria of $1 \times 10^{-5}$ eV are met. For each system, the unit cell is optimized to obtain the lattice parameters corresponding to the lowest energy. Uniaxial tensile strain along the direction of nanoribbon (or nanotube axis) is applied. Effects of transverse electric field along the lateral direction of the nanoribbon (or nanotube) are investigated. The uniform electric field is handled in VASP by adding an artificial dipole sheet in the supercell.[48] The geometries are kept fixed when applying the external electric field to avoid the geometric contribution to the electronic structures. Since DFT/PBE method tends to underestimate bandgap of semiconductors, the screen hybrid HSE06 method is also used to examine the band structures.[49]

**Acknowledgments.** The USTC group is supported by the National Basic Research Programs of China (Nos. 2011CB921400, 2012CB 922001), NSFC (Grant Nos. 21121003, 11004180, 51172223), One Hundred Person Project of CAS, Strategic Priority Research Program of CAS (XDB01020300), Shanghai Supercomputer Center, and Hefei Supercomputer Center. UNL group is supported by ARL (Grant No. W911NF1020099), NSF (Grant No. DMR-0820521), Nebraska Center for Energy Sciences Research and UNL Holland Computing Center, and a grant from USTC for (1000plan) Qianren-B summer research.

*Supporting Information Available*: The cell parameter $a_1$ and $a_2$ of multilayer phosphorene and the interlayer spacing between two adjacent phosphorene layers $d_{int}$, computed band structures (PBE) of zigzag and armchair PNRs, band structures (PBE) of 8-armchair PNR at the strain of -4% and 4%, band structures (PBE) and partial charge density of 8-armchair PNR at the external electric field of 0 V/Å and 0.1 V/Å, cohesive energy per atom of PNRs as a function of the ribbon width ($7 \leq N \leq 12$), band structures (PBE) of armchair PNT (0, 8) at the strain of -4% and 4%, and band structures (PBE) of zigzag PNT (12, 0) at the strain of -4% and 4%, geometric structure of multilayer phosphorene, computed band structures (based on HSE06) of multilayer phosphorene, and partial density of states (PDOS) of $MoS_2$/phosporene and $WS_2$/phosporene heterobilayers. These materials are available free of charge *via* the Internet at http://pubs.acs.org.

TOC Graphic

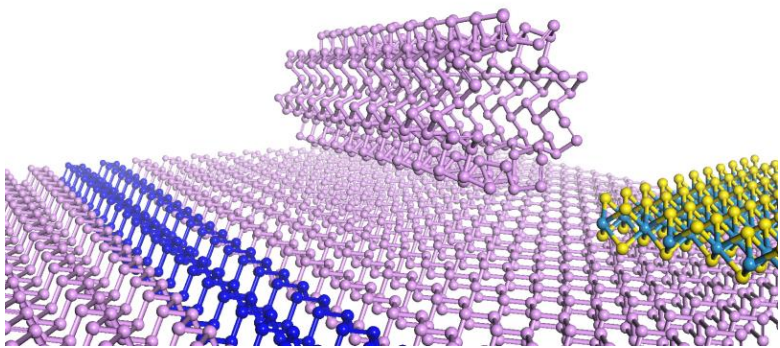